\documentclass{ws-p10x7}

\begin{document}

\title{
Centrality dependence of $K^+$ produced in Pb+Pb collisions
 at 158 GeV per nucleon}

\author{ Sonja Kabana for the NA52 collaboration:}

\author{
 R~Arsenescu, H~P~Beck, K~Borer, S~Kabana, R~Klingenberg, G~Lehmann,
R~Mommsen, U~Moser,  K~Pretzl, J~Schacher, R~Spiwoks, M~Weber
}
\address{Laboratory for High Energy Physics, University of Bern,
    Sidlerstrasse 5, CH-3012 Bern, Switzerland
(E-mail: sonja.kabana@cern.ch)
}

\author{
K~Elsener, K~D~Lohmann
}
\address{CERN, CH-1211 Geneva 23, Switzerland}

\author{
C~Baglin, A~Bussi\`ere, J~P~Guillaud
}
\address{ CNRS-IN2P3, LAPP Annecy, F-74941 Annecy-le-Vieux, France}

\author{
T~Lind\'en, J~Tuominiemi
}
\address{Dept. of Physics and Helsinki Institute of Physics, 
 University of Helsinki,
    PO Box 9, FIN-00014 Helsinki, Finland}

\author{
Ph~Gorodetzky
}
\address{PCC-College de France, 11 place Marcelin Berthelot, 75005 Paris, France}

\twocolumn[\maketitle\abstract{
\noindent
The NA52 collaboration searches for a
 discontinuous behaviour of charged kaons produced
in Pb+Pb collisions at 158 A GeV as a function of the impact parameter,
which could reveal a hadron to quark-gluon plasma (QGP) phase transition.
The $K+$ yield is found to grow
 proportional to the number of participating ('wounded') nucleons
N, above N=100.
Previous NA52 data agree with the above finding
and show a discontinuous behaviour in the kaon
centrality dependence near N=100, 
marking the onset of strangeness enhancement -over e.g. p+A data
at the same $\sqrt{s}$-
 in a chemically equilibrated phase.
}]

\section{Introduction}

\noindent
The hadron to 
quark-gluon plasma phase transition predicted by QCD \cite{qcd}
 may occur and manifest itself in ultrarelativistic heavy ion collisions
 through discontinuities in the 
energy density dependence of relevant observables.
A major example of such a discontinuity 
is seen in the $J/\Psi/DY$ ratio \cite{na50_jpsi}.
The NA52 collaboration searches for discontinuities in strangeness
production measuring charged kaons 
as a function of the impact parameter.
 Results from the 1995 NA52 run are published in \cite{na52_centr}.
We report here on new preliminary results
from the 1998 run of the NA52 experiment, 
on $K^+$ at rapidity 4.1 and transverse momentum near 0
produced in Pb+Pb collisions at 158 A GeV \cite{my_sqm2000_1}.
In this run
a new electromagnetic lead/quartz fiber  calorimeter (QFC) 
with improved acceptance and resolution \cite{qfc}
was used.

\section{Results and discussion}

\noindent
For the kaon measurement 
we modified the 1998 set up of NA52 \cite{michele_sqm2000}
 by placing the target 0.6 m upstream of the calorimeter.
The results 
have been corrected for empty target contributions.
The number of participant nucleons N has been estimated from
the energy measured with the calorimeter (figure 1) in the way
described in \cite{na52_centr}. 
Particle identification is described in \cite{na52_centr}
and references there.
\begin{figure}
\epsfxsize190pt
\figurebox{}{}{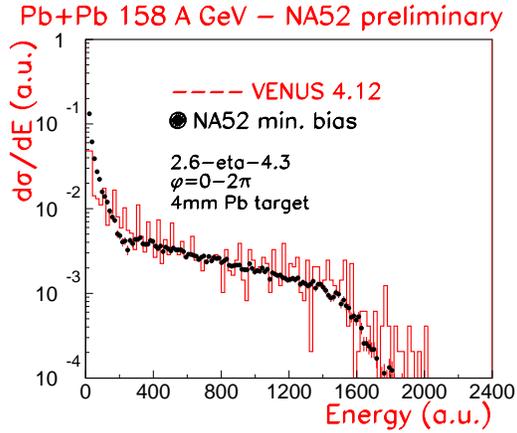}
\caption{
Preliminary energy distribution in arbitrary units 
in minimum bias Pb+Pb collisions at 158 A GeV,
 from the 1998 NA52 run.
}
\label{fig:e}
\end{figure}
\noindent 
The positive kaon yield divided by N
is independent of N, for N $>$ 100 (figure 2)
in agreement with previous NA52 results \cite{na52_centr}.
Assuming that N is proportional to the volume of the particle source,
figure 2 
 shows that the kaon number density 
exhibit a discontinuity, saturating  above N=100.
\begin{figure}
\epsfxsize190pt
\figurebox{}{}{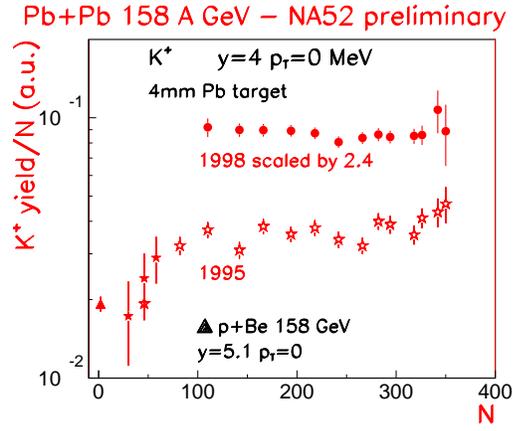}
\caption{
Preliminary 
$K^+$ yield in arbitrary units per participant nucleon N, as a function  of
N from Pb+Pb collisions at 158 A GeV, measured in the 1998 
NA52 run.
For comparison the 1995 NA52 data \protect\cite{na52_centr} are also
shown. 
The 1998 data are scaled by 2.4 with respect to the 1995 data.
}
\label{fig:k}
\end{figure}

This  indicates a transition to a phase characterized by a high degree of 
chemical equilibrium and enhancement \cite{na52_centr}
of kaons from the point N=100 on, corresponding to energy density
 $\epsilon$ $\sim$ 1.3 GeV/fm$^3$ \cite{hepph_0004138,my_sqm2000_2,2talk}, near
the critical energy density $\epsilon_c$$\sim$1-2 GeV/fm$^3$
 \cite{qcd,satz_review}.
This change marks the onset of strangeness enhancement seen in kaons
 in an equilibrated phase,
which may be suggestive for a QCD phase transition,
depending on the simultaneous appearance of thresholds in other signatures 
like e.g. the $J/\Psi$ suppression 
 at the relevant $\epsilon$ values and their theoretical understanding
  \cite{satz_review,na50_jpsi,hepph_0004138,pbm}.

\section*{Acknowledgments}
We thank Prof.~P. Minkowski for fruitfull discussions and 
the Schweizerischer Nationalfonds for their support.

\end{document}